\documentclass[12pt]{article}
\usepackage{amssymb, amsmath, amsfonts, eurosym, geometry, graphicx, color, setspace, sectsty, comment, footmisc, caption, natbib, pdflscape, array, hyperref, float, verbatim, listings, xcolor, commath, enumitem, xfrac, pgfplots, mathtools, tikz, multirow, diagbox, chngpage, booktabs, tabularx, dirtytalk, indentfirst, comment, setspace, subcaption, rotating, longtable, makecell, ltablex}

\usepackage[utf8]{inputenc}
\usepackage[affil-it]{authblk}

\spacing{1.25}

\begin{document}

\bibliographystyle{chicago} 
\setcitestyle{round} 

\title{Modern Portfolio Diversification with Arte-Blue Chip Index \\
}

\author[1]{Simon Levy\thanks{Corresponding author. E-mail address: \url{simon@levysimon.com}}}
\author[1]{Maxime L. D. Nicolas}
\affil[1]{\small{\textit{Institute of Finance \& Technology, University College London, UK}}}

\date{}

\maketitle


\vspace{-1.cm}

\begin{abstract}
\linespread{1}\selectfont

This paper presents a novel approach to evaluating blue-chip art as a viable asset class for portfolio diversification. We present the Arte-Blue Chip Index, an index that tracks 100 top-performing artists based on 81,891 public transactions from 157 artists across 584 auction houses over the period 1990 to 2024. By comparing blue-chip art price trends with stock market fluctuations, our index provides insights into the risk and return profile of blue-chip art investments. Our analysis demonstrates that a 20\% allocation of blue-chip art in a diversified portfolio enhances risk-adjusted returns by around 20\%, while maintaining volatility levels similar to the S\&P 500.
\noindent

\\\

\end{abstract}

\thispagestyle{empty}
\clearpage

\newpage

\section{Introduction}

In financial markets, blue-chip stocks are known for their reliability and long-standing performance. Similarly, in the art world, blue-chip art refers to works by significant artists such as Picasso, Warhol, and Richter, whose pieces consistently retain or increase in value. These artists have established auction histories, with individual works often selling for over \$500,000. 

The stability and resilience of blue-chip art make it appealing, as it often retains or increases its value during economic downturns \citep{candela1997price}. The low correlation with traditional financial assets, such as stocks and bonds, makes blue-chip art an attractive option for portfolio diversification \citep{worthington2004art}. Recent advancements in digital platforms and online auctions have expanded access to blue-chip art, allowing a broader range of investors to engage in the market \citep{thompson201212}. This has strengthened its appeal not only for aesthetic appreciation but also as a viable asset class for diversification. However, despite its importance, there is still a lack of focused academic research on the specific role of blue-chip art in investment diversification, as most existing literature addresses the broader art market without isolating this segment.

Previous research has highlighted lower returns and higher volatility for the broader art market \citep{baumol1986unnatural, frey1989muses, mei2002art, goetzmann1993accounting, renneboog2013buying, david2013art}. Despite this, blue-chip art, with its stable long-term value and strong market presence, remains underexplored in this area. Moreover, the lack of specific indices to capture blue-chip art performance further exacerbates this gap, leaving investors without a reliable measure of performance for this niche segment of the art market.

To fill this gap, we developed the Arte-Blue Chip Index, which tracks the performance of 100 top-performing blue-chip artists over a 34-year period (1990-2024). This index is built on 81,891 public transactions from 157 artists across 584 auction houses. The index focuses on each artist’s preferred and rarest mediums—primarily paintings and sculptures—while excluding lesser-valued mediums such as works on paper or editions. The index is rebalanced annually to ensure that only the top-performing artists are included based on their auction performance each year. Additionally, we demonstrate the potential of blue-chip art as a stable diversifier in investment portfolios. To address the inherent illiquidity of art, we simulate real-world investment scenarios with flexible entry and exit windows, aligning art investments more closely with public market dynamics.

Our results demonstrate that incorporating blue-chip art into an investment portfolio improves risk-adjusted returns without significantly increasing portfolio volatility. Specifically, a 20\% allocation to blue-chip art in a diversified portfolio enhanced returns by around 20\%, while maintaining volatility levels comparable to the S\&P 500. Additionally, our simulations show that flexible entry and exit points for art investments, when combined with the Arte-Blue Chip Index, provide liquidity solutions that better align art investments with traditional market dynamics.

The estimation of art investment performance has evolved significantly. Early studies, such as those by \cite{baumol1986unnatural} and \cite{frey1989muses}, used geometric mean returns to estimate art investment performance based on 640 transactions spanning from 1652 to 1961. However, these studies did not account for variations in the quality of artworks. To address this limitation, repeat-sales regression (RSR) techniques were introduced, which use the prices of artworks sold at least twice to control for quality differences. \cite{goetzmann1993accounting} employed this method to construct an art return index using transaction prices from 1715 to 1986, improving the accuracy of performance measures by focusing on repeat sales.

Subsequent work by \cite{mei2002art} expanded on this methodology by constructing a new repeat-sales dataset from auction records at the New York Public Library and the Watson Library at the Metropolitan Museum of Art, compiling 4,896 price pairs covering the period from 1875 to 2000. Further refinements came from \cite{pesando2008auction}, who analyzed 80,214 repeat sales of modern prints sold at global auctions between 1977 and 2004. They found a modest real return of 1.51\% on a diversified portfolio of modern prints, addressing various factors impacting the performance of such investments. \cite{renneboog2013buying} broadened the scope by analyzing over one million auction transactions from the Art Sales Index database. They concluded that art investments appreciated at a moderate 3.97\% annually in real U.S. dollar terms between 1957 and 2007, providing the most extensive dataset for assessing long-term trends in the art market.

Regarding diversification potential, several studies highlight both the limitations and opportunities of art as an investment. \cite{baumol1986unnatural} found that art generally underperforms compared to stocks and bonds, offering lower returns and higher volatility, particularly when accounting for illiquidity and high transaction costs. More recent studies, such as the Mei Moses Art Index \citep{mei2002art}, emphasize art's potential for diversification, though they confirm that it often underperforms equities.

\cite{renneboog2013buying, korteweg2016does} also found that art can provide diversification benefits due to its low correlation with traditional assets, but they acknowledge the volatility and high costs associated with it. Additionally, \cite{masset2018rationality} compared art with collectibles and concluded that high-value art often yields lower returns, while also overlooking the non-financial rewards of ownership.

Despite these advancements, most studies neglect the specific role of blue-chip art in portfolio diversification, focusing instead on broader segments of the art market like modern prints or contemporary art. This paper fills that gap by introducing the Arte-Blue Chip Index, a data-driven tool that offers a reliable measure of blue-chip art performance. We demonstrate that blue-chip art provides not only aesthetic value but also serves as a stable, long-term investment with clear financial benefits.

The remainder of this paper is organized as follows: Section \ref{sec:data-method} outlines the data and methods used to construct the Arte-Blue Chip Index. Section \ref{sec:results} presents the empirical results of portfolio optimization and performance simulations. Finally, Section \ref{sec:conclusion} offers conclusions and discusses the implications of our findings for investors and the art market.

\section{Data \& Methods} \label{sec:data-method}

We collected data from 584 major auction platforms, including Artsy and Artprice, to ensure comprehensive coverage of market transactions. Our dataset was specifically filtered to include only high-value mediums—primarily paintings and sculptures—traditionally more valuable in the auction market. To maintain consistency in measuring the financial performance of blue-chip artworks, we excluded lower-value mediums such as prints and works on paper. To qualify for inclusion, artists needed at least a decade of auction history, with consistent sales of their works exceeding \$500,000 in average. This filtering process produced a final dataset of 81,891 public auction transactions from 157 selected artists, spanning from 1990 to 2024. Each transaction record contained detailed information, including the artwork’s title, medium, artist’s name, auction house, sale date, dimensions (height, width, and area), sale price, pre-sale estimate, and performance relative to the pre-sale estimate.

To standardize prices based on the size of the artwork, we calculated the price per unit area using the provided dimensions. For each artist, we then computed the average normalized price per year. These normalized prices were then combined with the artist's average artwork areas to estimate actual prices for each year. To assess artist performance, we calculated the average actual prices at both the start and end of the period. From these, we derived performance metrics, including yearly returns and cash-on-cash returns. Additionally, the Internal Rate of Return (IRR) was calculated based on the average transaction prices over the period, and the Multiple of Invested Capital (MOIC) was used to measure the total return on investment.

Performance metrics for the artists are detailed in Table \ref{tab:irr_moic} for the investment period from 2005 to 2015, and in Table \ref{tab:irr_moic_2012_2015} for the period from 2012 to 2015. To prevent selection bias, we rebalance each year the index to retain up to 100 top-performing artists. For each selected year, artist weights were calculated based on their average prices relative to the total average prices of all artists in the index. For example, Pablo Picasso had the highest weight in 2022 at 13.58\%, followed by Andy Warhol at 11.76\%. Table \ref{tab:artist_weights} provides an example of artist weights for 2022.

For comparison, we aligned our index with the methodology of Artprice's blue-chip art index, which identifies the 100 top-performing artists over the previous five years, with annual updates. The aggregated optimized performance of our selected artists was compared to their approach. 

\section{Results} \label{sec:results}

We measured the performance of the art index based on a 680-day moving average which corresponds to the average period in such investment context.\footnote{This is usually the proposed investment period by platforms such as Matis.Club and Masterworks.com that allow fractional investment of high-value art pieces.} We then simulated an investment portfolio with an initial 20/80 allocation between the Arte-Blue Chip Index and the S\&P 500.

\begin{figure}[H]
    \centering   
    \caption{Cumulative Returns from 1990 to 2024}
    \label{fig:index}
    \includegraphics[width=0.8\linewidth]{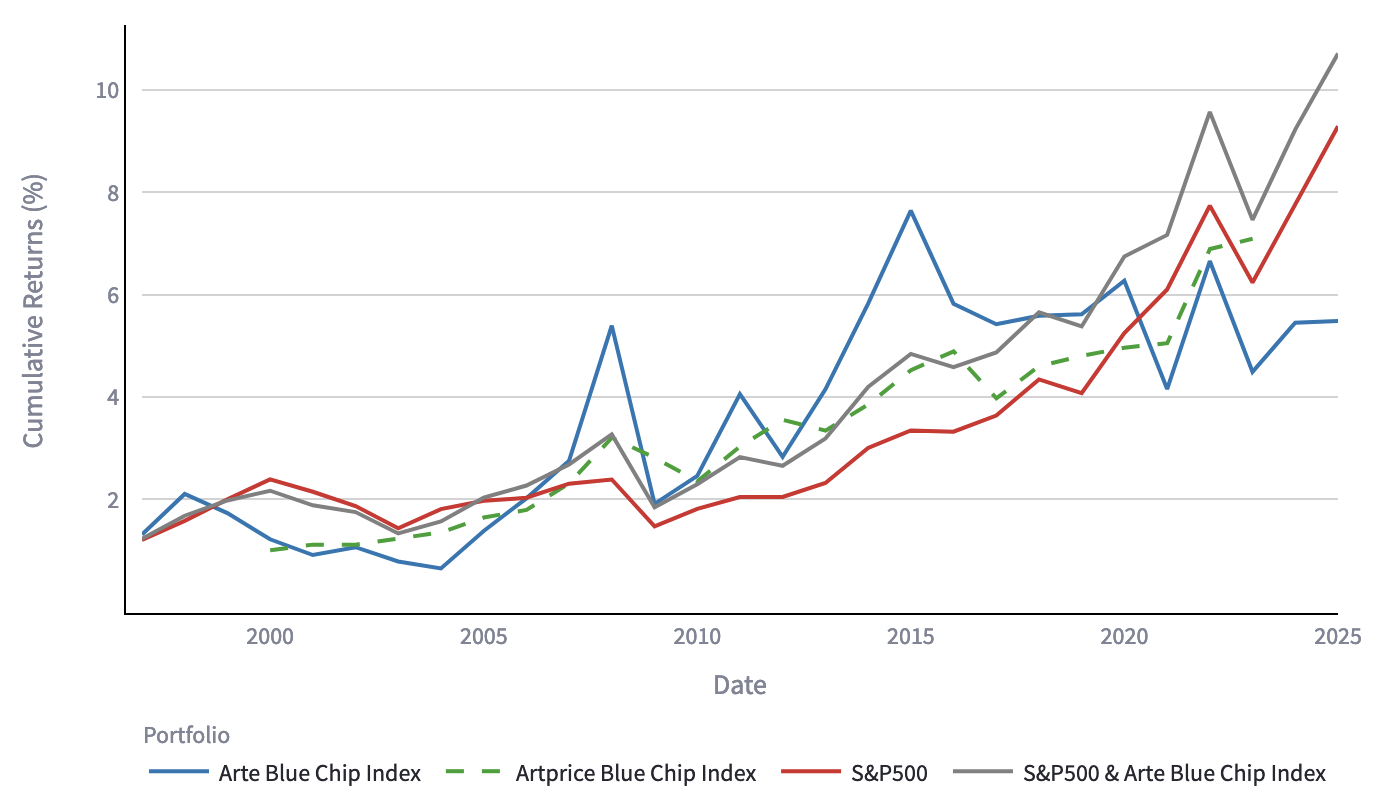}
\resizebox{\textwidth}{!}{%
    \begin{tabular}{p{\textwidth}}
    \small{ \textbf{Notes:} This figure displays the cumulative returns for an 80/20 S\&P 500/Arte allocation, using a 680-day rolling window.
    }
    \end{tabular}}
\end{figure}

The results are shown in Figure \ref{fig:index}. The inclusion of blue-chip art significantly improved the overall portfolio performance, achieving a higher cumulative return (922.49\%) compared to both the S\&P 500 (774.41\%) and the Arte-Blue Chip Index alone (544.99\%). This performance was achieved with relatively low volatility (66.27\%) compared to the Arte-Blue Chip Index (145.68\%) and only slightly higher than the S\&P 500 (60.62\%). Additionally, the portfolio’s Sharpe Ratio of 1.49 demonstrates better risk-adjusted returns, indicating more efficient balancing of risk and reward compared to both the S\&P 500 and the Arte-Blue Chip Index alone. For the period between 2007 and 2024, we observed an average correlation of 0.51 between the Arte-Blue Chip Index and the S\&P 500, ranging from 0.12 in 2007 to 0.75 in 2019.

\begin{figure}[H]
    \centering
    \caption{Annual Returns from 1990 to 2024} \label{fig:simulation}
    \includegraphics[width=0.8\linewidth]{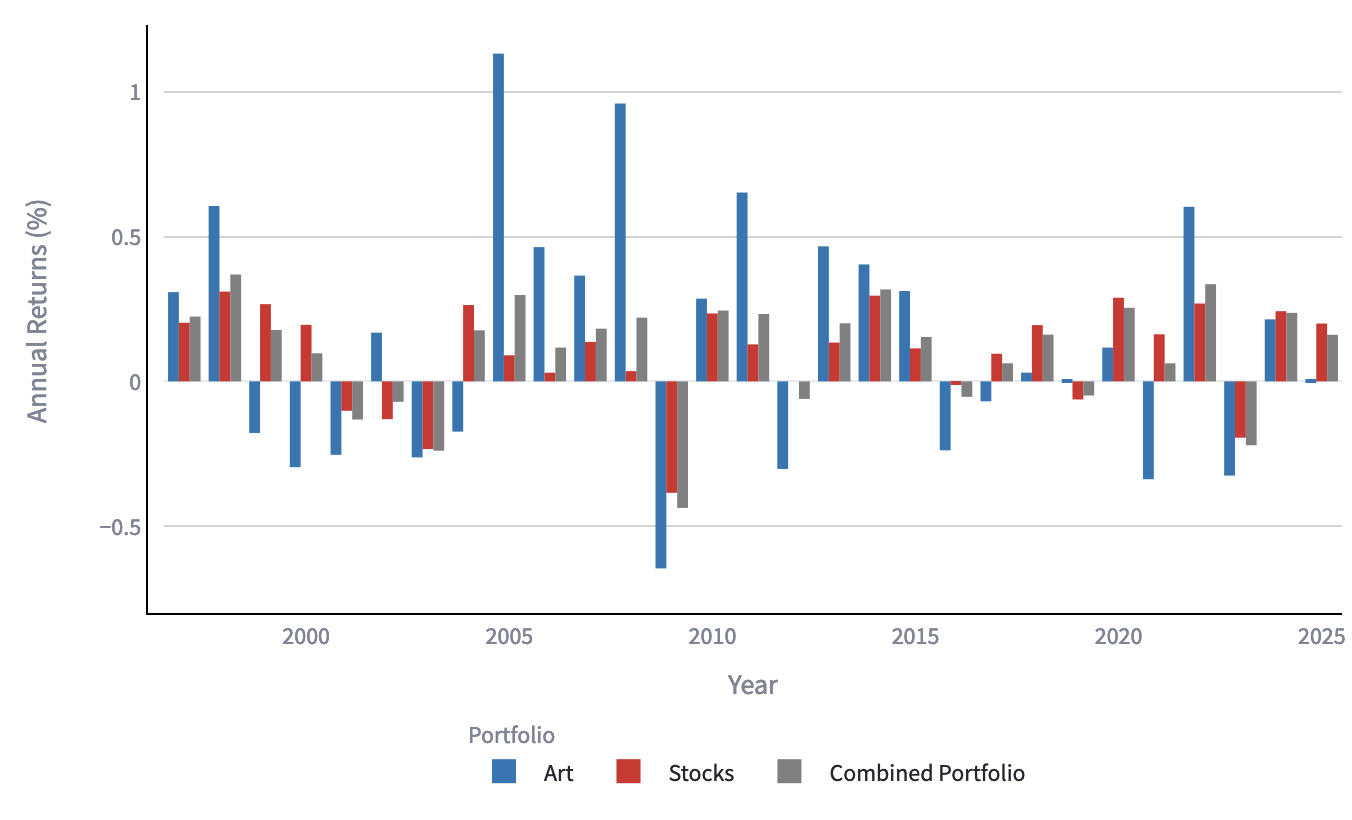}
\resizebox{\textwidth}{!}{%
    \begin{tabular}{p{\textwidth}}
    \small{ \textbf{Notes:} This figure displays the annual returns with an 80/20 S\&P 500/Arte-Blue Chip Index allocation, using a 680-day rolling window.
    }
    \end{tabular}}
\end{figure}

\begin{figure}[H]
    \centering   
    \caption{Efficient Frontier}
    \includegraphics[width=0.8\linewidth]{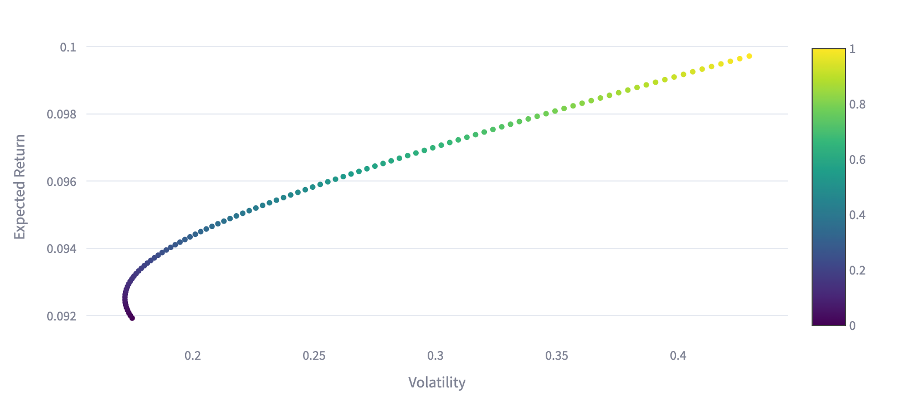}
\resizebox{\textwidth}{!}{%
    \begin{tabular}{p{\textwidth}}
    \small{ \textbf{Notes:} This figure presents the efficient frontier for various portfolio allocations, based on the risk-free rate and the rolling average for the Arte-Blue Chip Index allocation.
    }
    \end{tabular}}
    \label{fig:6}
\end{figure}

In Figure \ref{fig:simulation}, we present the average annual returns and observe that with an 80/20 allocation between the S\&P 500 and the Arte-Blue Chip Index, increase annual returns. Additionally, the efficient frontier shown in Figure \ref{fig:6} indicates that the optimal risk-adjusted return, with the lowest volatility, is achieved with approximately a 10\% allocation to the Arte-Blue Chip Index. Although a 20\% allocation yields higher returns, it also increases volatility, making the 10\% allocation a more suitable choice for balancing risk and return.

\section{Conclusion} \label{sec:conclusion}

Incorporating blue-chip art into investment portfolios offers notable diversification and risk reduction advantages. Our analysis demonstrates that blue-chip art can enhance risk-adjusted returns. Despite challenges such as high transaction costs and market illiquidity, integrating art into a portfolio, as illustrated by our 80/20 with the S\&P500 allocation model, has shown superior performance compared to traditional and art-only indices. The combination of high cumulative returns and favourable risk metrics underscores the potential of blue-chip art to contribute positively to diversified investment strategies.
Investors looking to diversify beyond traditional assets may find blue-chip art to be a valuable alternative.


\bibliography{biblio}

\appendix

\begin{small}
\begin{longtable}{|l|r|r|r|r|}
\caption{Artists performances 2005 – 2015} \label{tab:irr_moic} \\
\hline
\textbf{Artist} & \textbf{\makecell{Avg Price \\ Initial (\$K)}} & \textbf{\makecell{Avg Price \\ Final (\$K)}} & \textbf{IRR (\%)} & \textbf{\makecell{Avg \\MOIC}}  \\
\hline
\endfirsthead

\caption{Artists performances 2005 – 2015 (continued)} \\
\hline
\textbf{Artist} & \textbf{\makecell{Avg Price \\ Initial (\$K)}} & \textbf{\makecell{Avg Price \\ Final (\$K)}} & \textbf{IRR (\%)} & \textbf{\makecell{Avg \\MOIC}}  \\
\hline
\endhead

\hline
\multicolumn{5}{r}{\textit{Continued on next page}} \\

\endfoot

\hline
\endlastfoot

Robert Ryman & 35.00 & 16,010.01 & 84.52\% & 457.43  \\
Lee Ufan & 1.11 & 354.62 & 77.98\% & 318.90  \\
Gunther Forg & 2.24 & 307.18 & 63.55\% & 136.90  \\
Salvador Dali & 9.12 & 1,212.45 & 63.07\% & 132.93  \\
Ernst Ludwig Kirchner & 37.74 & 2,912.13 & 54.43\% & 77.15  \\
Robert Rauschenberg & 70.00 & 3,090.95 & 46.05\% & 44.16  \\
Cy Twombly & 388.24 & 12,600.98 & 41.62\% & 32.46  \\
Rudolf Stingel & 27.47 & 798.94 & 40.07\% & 29.08  \\
Norman Rockwell & 48.00 & 1,360.27 & 39.71\% & 28.34  \\
Christopher Wool & 203.66 & 5,231.60 & 38.35\% & 25.69  \\
Peter Doig & 324.94 & 7,677.29 & 37.20\% & 23.63  \\
Georges Braque & 84.10 & 1,933.57 & 36.82\% & 22.99  \\
Tracey Emin & 46.67 & 968.86 & 35.43\% & 20.76  \\
Anselm Kiefer & 33.03 & 678.06 & 35.28\% & 20.53  \\
Paul Signac & 129.76 & 2,616.04 & 35.04\% & 20.16  \\
René Magritte & 223.12 & 3,977.27 & 33.38\% & 17.83  \\
David Hockney & 500.00 & 8,357.04 & 32.53\% & 16.71  \\
Brice Marden & 214.80 & 2,283.07 & 26.66\% & 10.63  \\
Joan Mitchell & 356.56 & 3,131.39 & 24.27\% & 8.78  \\
Robert Longo & 26.04 & 220.93 & 23.84\% & 8.48  \\
Lucio Fontana & 437.21 & 3,660.26 & 23.67\% & 8.37  \\
Jean Michel Basquiat & 952.80 & 7,431.25 & 22.80\% & 7.80  \\
Yayoi Kusama & 93.59 & 721.69 & 22.66\% & 7.71  \\
Andy Warhol & 881.19 & 6,598.95 & 22.30\% & 7.49  \\
Henry Moore & 210.18 & 1,553.16 & 22.14\% & 7.39  \\
Francis Picabia & 94.81 & 640.62 & 21.05\% & 6.76  \\
Henri Matisse & 659.17 & 4,317.97 & 20.68\% & 6.55  \\
Wayne Thiebaud & 423.48 & 2,769.90 & 20.66\% & 6.54  \\
Chu Teh Chun & 165.47 & 932.77 & 18.88\% & 5.64  \\
Wassily Kandinsky & 402.56 & 2,194.50 & 18.48\% & 5.45  \\
Marc Chagall & 318.48 & 1,530.77 & 17.00\% & 4.81  \\
Zao Wou Ki & 457.39 & 2,142.84 & 16.70\% & 4.68  \\
Helen Frankenthaler & 150.00 & 682.04 & 16.35\% & 4.55  \\
Dong Qichang & 76.20 & 344.69 & 16.29\% & 4.52  \\
Jim Dine & 10.16 & 45.83 & 16.26\% & 4.51  \\
Frank Stella & 3,181.58 & 14,355.24 & 16.26\% & 4.51  \\
Sam Gilliam & 8.15 & 36.37 & 16.14\% & 4.46  \\
Mark Rothko & 6,049.04 & 25,372.15 & 15.42\% & 4.19  \\
Alexander Calder & 2,955.05 & 12,277.77 & 15.31\% & 4.15  \\
Howard Hodgkin & 185.93 & 759.92 & 15.12\% & 4.09  \\
Milton Avery & 134.75 & 548.43 & 15.07\% & 4.07  \\
Pablo Picasso & 3,034.01 & 12,319.69 & 15.04\% & 4.06  \\
Wu Hufan & 40.06 & 160.92 & 14.92\% & 4.02  \\
Ed Ruscha & 321.63 & 1,256.62 & 14.60\% & 3.91  \\
George Condo & 43.65 & 156.88 & 13.65\% & 3.59  \\
Camille Pissarro & 63.59 & 213.10 & 12.86\% & 3.35  \\
Albert Oehlen & 177.50 & 587.20 & 12.71\% & 3.31  \\
Gerhard Richter & 567.96 & 1,855.02 & 12.56\% & 3.27  \\
Claude Monet & 2,384.05 & 7,660.70 & 12.38\% & 3.21  \\
Max Ernst & 385.77 & 1,155.40 & 11.59\% & 2.99  \\
Antoni Tapies & 52.44 & 155.99 & 11.52\% & 2.97  \\
Auguste Rodin & 88.72 & 260.17 & 11.36\% & 2.93  \\
Lynn Chadwick & 57.89 & 169.04 & 11.31\% & 2.92  \\
Kees van Dongen & 382.55 & 1,012.93 & 10.23\% & 2.65  \\
Georgia OKeefe & 471.12 & 1,119.61 & 9.04\% & 2.38  \\
Jean Dubuffet & 398.78 & 935.87 & 8.91\% & 2.35  \\
Takashi Murakami & 156.38 & 352.46 & 8.47\% & 2.25  \\
Yves Klein & 615.73 & 1,379.62 & 8.40\% & 2.24  \\
Richard Prince & 735.89 & 1,638.07 & 8.33\% & 2.23  \\
George Mathieu & 36.40 & 77.76 & 7.89\% & 2.14  \\
Alfred Sisley & 934.39 & 1,865.62 & 7.16\% & 2.00  \\
Tsuguharu Foujita & 202.97 & 403.48 & 7.11\% & 1.99  \\
Bernard Buffet & 79.71 & 157.51 & 7.05\% & 1.98  \\
Jean-Paul Riopelle & 162.14 & 317.62 & 6.95\% & 1.96  \\
Julian Shnabel & 110.54 & 215.59 & 6.91\% & 1.95  \\
Hans Hofman & 487.36 & 919.58 & 6.56\% & 1.89  \\
Anish Kapoor & 240.00 & 435.74 & 6.15\% & 1.82  \\
Robert Mapplethorpe & 1,637.26 & 2,965.52 & 6.12\% & 1.81  \\
Josef Albers & 175.35 & 317.28 & 6.11\% & 1.81  \\
Sam Francis & 66.47 & 119.85 & 6.07\% & 1.80  \\
Willem de Kooning & 3,734.12 & 6,664.69 & 5.96\% & 1.78  \\
Tamara de Lempicka & 196.51 & 342.58 & 5.72\% & 1.74  \\
Egon Schiele & 250.01 & 435.47 & 5.71\% & 1.74  \\
Joan Miro & 2,349.01 & 4,083.45 & 5.69\% & 1.74  \\
Roberto Matta & 146.72 & 251.87 & 5.55\% & 1.72  \\
Louis Nevelson & 40.28 & 66.72 & 5.18\% & 1.66  \\
George Baselitz & 637.42 & 1,044.77 & 5.07\% & 1.64  \\
Franz Kline & 2,699.63 & 4,363.69 & 4.92\% & 1.62  \\
Hans Hartung & 99.87 & 154.91 & 4.49\% & 1.55  \\
Antony Gormley & 531.39 & 814.98 & 4.37\% & 1.53  \\
Le Pho & 25.77 & 38.32 & 4.05\% & 1.49  \\
Karel Appel & 50.57 & 72.22 & 3.63\% & 1.43  \\
Tom Wesselman & 697.36 & 926.30 & 2.88\% & 1.33  \\
Camille Pissaro & 842.74 & 1,083.11 & 2.54\% & 1.29  \\
Sigmar Polke & 296.28 & 371.39 & 2.29\% & 1.25  \\
Leonor Fini & 49.51 & 62.05 & 2.28\% & 1.25  \\
Nicolas De Stael & 382.42 & 459.10 & 1.84\% & 1.20  \\
Maurice de Vlaminck & 71.29 & 84.72 & 1.74\% & 1.19  \\
Agnes Martin & 1,300.31 & 1,539.00 & 1.70\% & 1.18  \\
Giorgio Morandi & 202.44 & 235.55 & 1.53\% & 1.16  \\
Eugene Boudin & 120.52 & 139.78 & 1.49\% & 1.16  \\
Wen Zhengming & 129.55 & 146.68 & 1.25\% & 1.13  \\
Damien Hirst & 284.87 & 322.33 & 1.24\% & 1.13  \\
Giorgio de Chirico & 162.93 & 183.27 & 1.18\% & 1.12  \\
Paul Cezanne & 7,136.93 & 7,959.67 & 1.10\% & 1.12  \\
Pierre-Auguste Renoir & 260.08 & 273.69 & 0.51\% & 1.05  \\
Cindy Sherman & 202.36 & 211.98 & 0.47\% & 1.05  \\
Raoul Dufy & 172.35 & 178.90 & 0.37\% & 1.04  \\
Keith Haring & 155.77 & 155.74 & -0.00\% & 1.00  \\
Serge Poliakoff & 186.75 & 182.23 & -0.24\% & 0.98  \\
Niki de Saint Phalle & 125.65 & 103.97 & -1.88\% & 0.83  \\
\end{longtable}
\end{small}

\begin{small}
\begin{longtable}{|l|r|r|r|r|}
\caption{Artists performances 2012 – 2015} \label{tab:irr_moic_2012_2015} \\
\hline
\textbf{Artist} & \textbf{\makecell{Avg Price \\ Initial (\$K)}} & \textbf{\makecell{Avg Price \\ Final (\$K)}} & \textbf{IRR (\%)} & \textbf{\makecell{Avg \\MOIC}}  \\
\hline
\endfirsthead

\caption{Artists performances 2012 – 2015 (continued)} \\
\hline
\textbf{Artist} & \textbf{\makecell{Avg Price \\ Initial (\$K)}} & \textbf{\makecell{Avg Price \\ Final (\$K)}} & \textbf{IRR (\%)} & \textbf{\makecell{Avg \\MOIC}}  \\
\hline
\endhead

\hline
\multicolumn{5}{r}{\textit{Continued on next page}} \\

\endfoot

\hline
\endlastfoot

Adrian Ghenie & 13.48 & 1,565.00 & 387.80\% & 116.07  \\
David Hockney & 86.85 & 8,357.04 & 358.24\% & 96.22  \\
Peter Doig & 113.35 & 7,677.29 & 307.63\% & 67.73  \\
Robert Rauschenberg & 114.58 & 3,090.95 & 199.91\% & 26.98  \\
Frank Stella & 535.77 & 14,355.24 & 199.23\% & 26.79  \\
Robert Mapplethorpe & 114.58 & 2,965.52 & 195.80\% & 25.88  \\
Tracey Emin & 95.88 & 968.86 & 116.19\% & 10.10  \\
Robert Ryman & 1,650.50 & 16,010.01 & 113.27\% & 9.70  \\
Brice Marden & 271.79 & 2,283.07 & 103.28\% & 8.40  \\
Henry Moore & 261.68 & 1,553.16 & 81.06\% & 5.94  \\
Lynn Chadwick & 30.30 & 169.04 & 77.36\% & 5.58  \\
Wayne Thiebaud & 500.11 & 2,769.90 & 76.93\% & 5.54  \\
Paul Cezanne & 1,538.50 & 7,959.67 & 72.95\% & 5.17  \\
Kazuo Shiraga & 214.55 & 1,006.27 & 67.39\% & 4.69  \\
Gunther Forg & 67.61 & 307.18 & 65.62\% & 4.54  \\
Auguste Rodin & 58.96 & 260.17 & 64.02\% & 4.41  \\
Kenny Scharf & 7.09 & 29.39 & 60.64\% & 4.15  \\
Ernst Ludwig Kirchner & 727.03 & 2,912.13 & 58.81\% & 4.01  \\
Julian Shnabel & 56.46 & 215.59 & 56.30\% & 3.82  \\
Andy Warhol & 2,184.76 & 6,598.95 & 44.55\% & 3.02  \\
Takashi Murakami & 118.97 & 352.46 & 43.63\% & 2.96  \\
Richard Prince & 553.22 & 1,638.07 & 43.60\% & 2.96  \\
Howard Hodgkin & 272.06 & 759.92 & 40.83\% & 2.79  \\
Lucio Fontana & 1,479.31 & 3,660.26 & 35.25\% & 2.47  \\
Tom Wesselman & 386.58 & 926.30 & 33.82\% & 2.40  \\
Norman Rockwell & 575.54 & 1,360.27 & 33.20\% & 2.36  \\
Sam Gilliam & 16.30 & 36.37 & 30.68\% & 2.23  \\
Dong Qichang & 156.95 & 344.69 & 29.98\% & 2.20  \\
Milton Avery & 250.78 & 548.43 & 29.80\% & 2.19  \\
Egon Schiele & 202.12 & 435.47 & 29.16\% & 2.15  \\
Francis Picabia & 299.15 & 640.62 & 28.90\% & 2.14  \\
Ed Ruscha & 599.00 & 1,256.62 & 28.01\% & 2.10  \\
Alexander Calder & 6,224.26 & 12,277.77 & 25.41\% & 1.97  \\
Antony Gormley & 417.33 & 814.98 & 24.99\% & 1.95  \\
Raoul Dufy & 92.03 & 178.90 & 24.80\% & 1.94  \\
Helen Frankenthaler & 353.07 & 682.04 & 24.54\% & 1.93  \\
Georges Braque & 1,009.21 & 1,933.57 & 24.20\% & 1.92  \\
Cy Twombly & 6,587.07 & 12,600.98 & 24.14\% & 1.91  \\
Alfred Sisley & 976.81 & 1,865.62 & 24.07\% & 1.91  \\
Kees van Dongen & 540.11 & 1,012.93 & 23.32\% & 1.88  \\
Cecily Brown & 601.40 & 1,110.94 & 22.70\% & 1.85  \\
Barbara Hepworth & 62.50 & 115.42 & 22.69\% & 1.85  \\
Albert Oehlen & 318.63 & 587.20 & 22.60\% & 1.84  \\
Willem de Kooning & 3,727.75 & 6,664.69 & 21.37\% & 1.79  \\
Christopher Wool & 2,980.26 & 5,231.60 & 20.63\% & 1.76  \\
George Baselitz & 595.63 & 1,044.77 & 20.60\% & 1.75  \\
Paul Gauguin & 1,762.50 & 3,050.19 & 20.06\% & 1.73  \\
Tsuguharu Foujita & 240.91 & 403.48 & 18.76\% & 1.67  \\
Damien Hirst & 194.30 & 322.33 & 18.38\% & 1.66  \\
Le Pho & 23.54 & 38.32 & 17.64\% & 1.63  \\
Camille Pissaro & 694.83 & 1,083.11 & 15.95\% & 1.56  \\
Marc Chagall & 1,002.01 & 1,530.77 & 15.17\% & 1.53  \\
Agnes Martin & 1,028.41 & 1,539.00 & 14.38\% & 1.50  \\
Hans Hartung & 109.10 & 154.91 & 12.40\% & 1.42  \\
Yayoi Kusama & 532.40 & 721.69 & 10.67\% & 1.36  \\
Pablo Picasso & 9,369.28 & 12,319.69 & 9.55\% & 1.31  \\
Roy Lichtenstein & 1,616.27 & 2,084.18 & 8.84\% & 1.29  \\
Bernard Buffet & 122.69 & 157.51 & 8.68\% & 1.28  \\
Roberto Matta & 200.15 & 251.87 & 7.96\% & 1.26  \\
Joan Mitchell & 2,555.37 & 3,131.39 & 7.01\% & 1.23  \\
Anish Kapoor & 356.53 & 435.74 & 6.92\% & 1.22  \\
Frank Auerbach & 316.97 & 363.35 & 4.66\% & 1.15  \\
Rudolf Stingel & 697.37 & 798.94 & 4.64\% & 1.15  \\
Zao Wou Ki & 1,873.00 & 2,142.84 & 4.59\% & 1.14  \\
Mark Rothko & 22,291.07 & 25,372.15 & 4.41\% & 1.14  \\
Jean-Paul Riopelle & 281.02 & 317.62 & 4.16\% & 1.13  \\
Claude Monet & 6,945.66 & 7,660.70 & 3.32\% & 1.10  \\
Josef Albers & 289.24 & 317.28 & 3.13\% & 1.10  \\
Paul Signac & 2389.70 & 2616.04 & 3.06\% & 1.09  \\
Chu Teh Chun & 874.58 & 932.77 & 2.17\% & 1.07  \\
Niki de Saint Phalle & 97.86 & 103.97 & 2.04\% & 1.06  \\
Robert Longo & 209.13 & 220.93 & 1.85\% & 1.06  \\
Ayako Rokkaku & 13.60 & 14.32 & 1.73\% & 1.05  \\
Jean Dubuffet & 918.28 & 935.87 & 0.63\% & 1.02  \\
Hans Hofman & 966.37 & 919.58 & -1.64\% & 0.95  \\
Wassily Kandinsky & 2,391.82 & 2,194.50 & -2.83\% & 0.92  \\
Wu Hufan & 177.64 & 160.92 & -3.24\% & 0.91  \\
Camille Pissarro & 239.30 & 213.10 & -3.79\% & 0.89  \\
Joan Miro & 4756.81 & 4,083.45 & -4.96\% & 0.86  \\
Leonor Fini & 74.35 & 62.05 & -5.85\% & 0.83  \\
Pierre Soulages & 657.68 & 545.22 & -6.06\% & 0.83  \\
Louis Nevelson & 80.77 & 66.72 & -6.17\% & 0.83  \\
John Chamberlain & 373.04 & 306.97 & -6.29\% & 0.82  \\
Robert Motherwell & 523.79 & 429.02 & -6.44\% & 0.82  \\
René Magritte & 4,913.41 & 3,977.27 & -6.80\% & 0.81  \\
Karel Appel & 89.40 & 72.22 & -6.86\% & 0.81  \\
Jim Dine & 57.18 & 45.83 & -7.11\% & 0.80  \\
Antoni Tapies & 195.02 & 155.99 & -7.17\% & 0.80  \\
Henri Matisse & 5,408.86 & 4,317.97 & -7.23\% & 0.80  \\
Elizabeth Peyton & 195.32 & 154.95 & -7.43\% & 0.79  \\
Eugene Boudin & 182.06 & 139.78 & -8.43\% & 0.77  \\
Pierre Bonnard & 736.01 & 536.82 & -9.98\% & 0.73  \\
Giorgio de Chirico & 252.39 & 183.27 & -10.12\% & 0.73  \\
Cindy Sherman & 305.89 & 211.98 & -11.51\% & 0.69  \\
Maurice Utrillo & 139.81 & 96.69 & -11.57\% & 0.69  \\
Max Ernst & 1672.18 & 1,155.40 & -11.59\% & 0.69  \\
Jean Michel Basquiat & 1,0835.59 & 7431.25 & -11.81\% & 0.69  \\
George Mathieu & 120.50 & 77.76 & -13.58\% & 0.65  \\
Fernand Leger & 1,083.30 & 669.83 & -14.81\% & 0.62  \\
Pierre-Auguste Renoir & 444.44 & 273.69 & -14.92\% & 0.62  \\
Bernar Venet & 200.19 & 119.12 & -15.89\% & 0.60  \\
\end{longtable}
\end{small}

\begin{small}
\begin{table}[ht]
\centering
\caption{Artist Weights Arte-Blue Chip 100 Index 2022}
\label{tab:artist_weights}
\begin{tabular}{|l|r|c|l|r|}
\cline{1-2} \cline{4-5}
\textbf{Artist} & \textbf{Weight (\%)} &  & \textbf{Artist} & \textbf{Weight (\%)} \\
\cline{1-2} \cline{4-5}
Aboudia & 0.29\% & & Jordan Kerwick & 0.20\% \\
Alexander Calder & 6.70\% & & Karel Appel & 0.28\% \\
Andy Warhol & 11.76\% & & Kenny Scharf & 0.61\% \\
Anselm Kiefer & 1.18\% & & Le Pho & 0.51\% \\
Arman & 0.03\% & & Louis Nevelson & 0.34\% \\
Auguste Rodin & 1.08\% & & Lucio Fontana & 6.07\% \\
Ayako Rokkaku & 0.90\% & & Lynn Chadwick & 0.48\% \\
Barbara Hepworth & 2.65\% & & Marc Chagall & 2.16\% \\
Bernard Buffet & 0.42\% & & Maurice Utrillo & 0.16\% \\
Camille Pissarro & 0.79\% & & Maurice de Vlaminck & 0.34\% \\
Christo & 0.25\% & & Max Ernst & 2.18\% \\
Cindy Sherman & 0.18\% & & Niki de Saint Phalle & 0.13\% \\
Damien Hirst & 1.03\% & & Pablo Picasso & 13.59\% \\
Elizabeth Peyton & 0.39\% & & Pierre Bonnard & 0.71\% \\
Eugene Boudin & 0.29\% & & Pierre-Auguste Renoir & 3.17\% \\
Francois-Xavier Lalanne & 0.65\% & & Raoul Dufy & 0.30\% \\
George Condo & 1.27\% & & Richard Serra & 0.48\% \\
George Mathieu & 0.74\% & & Roberto Matta & 0.37\% \\
Gerhard Richter & 9.19\% & & Sam Gilliam & 0.03\% \\
Giorgio de Chirico & 0.31\% & & Shara Hughes & 2.23\% \\
Gunther Forg & 0.78\% & & Tom Wesselman & 0.20\% \\
Hans Hartung & 0.17\% & & Yayoi Kusama & 3.22\% \\
Henry Moore & 2.53\% & & Yves Klein  & 3.59\% \\
Jean Dubuffet & 3.02\% & & Zao Wou Ki & 11.13\% \\
\cline{4-5}
Jean Michel Basquiat & 0.29\% & \multicolumn{1}{r}{} & \multicolumn{1}{r}{}\\ 
\cline{1-2} 
\end{tabular}
\end{table}
\end{small}

\end{document}